\documentclass[10pt,prc,aps]{revtex4}   \usepackage{graphicx}
\tighten
\begin{document}
\draft
\title{                                                     
Short-range correlations in isospin symmetric and asymmetric nuclear matter: \\
a microscopic perspective. 
 }              
\author{            
Francesca Sammarruca      }                                                                                
\affiliation{ Physics Department, University of Idaho, Moscow, ID 83844-0903, U.S.A. 
}
\date{\today} 
\begin{abstract}
Short-range correlations in nuclear and neutron matter are examined through the properties 
of the correlated wave function obtained by solving the Bethe-Goldstone equation. 
Tensor correlations are explored through the dominant tensor-driven transition 
and central correlations through the singlet and triplet $S$ waves.                
Predictions from a popular meson-theoretic nucleon-nucleon potential employed in the Dirac-Brueckner-Hartree-Fock       
approach are compared with those from two- and three-body high-quality chiral interactions in Brueckner $G$-matrix calculations. 
Short-range correlations in symmetric matter are remarkably stronger than in neutron matter.
It is found that short-range correlations are very model dependent and have a large impact on the symmetry energy above normal density. 
\end{abstract}
\maketitle

\section{Introduction} 
\label{Intro} 

Correlations in nuclear matter and nuclei carry important information about the 
underlying nuclear forces and their behavior in the medium. 
Since the early Brueckner nuclear matter calculations \cite{HT70}, it has been customary to associate the 
correlated two-body wave functions to the strength of the nucleon-nucleon (NN) potential. When this is done 
in a particular channel, one can extract information about specific components of the force.
 For instance, the $^3S_1-^3D_1$ channel will reveal tensor correlations, which 
have traditionally attracted particular attention,
 since the model dependence among predictions from different NN potentials resides mostly    
in the strength of their respective tensor force and its off-shell behavior. 

Today, nuclear interactions have reached a much higher level of sophistication. Furthermore, the impact 
of three-body forces, which generate additional tensor force, is                                   
a central question in contemporary nuclear physics, and should be addressed     
in any approach that wishes to be fundamental. 

On the experimental side, measurements at high-momentum transfer have detected remarkable differences
between correlations in $pn$ pairs, on the one hand, and $pp/nn$ pairs, on the other \cite{src,Pia+,CLAS,HallA}. Protons struck from the nucleus with initial momentum between the 
nucleus Fermi momentum and approximately 600 MeV/c, were found to emerge from a short-range correlated $pn$ 
pair 92\% of the times, whereas $pp/nn$ correlations were highly suppressed, contributing only 4\% of the high-momentum 
part of the distribution. Recalling that the tensor interaction impacts mostly the $np$ channel, and that the           
momentum region under consideration is tensor dominated, it is natural to conclude that one is looking 
at the effects of tensor correlations.                                                                  

Moreover, the tensor force plays a 
chief role in building up the symmetry energy (see Ref.~\cite{FS11} and references therein), which is the main 
mechanism in the formation of neutron skins as well as other systems/phenomena including radii
of compact stars. 

In summary, investigations of short-range correlations (of tensor nature, in particular) are of 
contemporary interest. Such investigations should be conducted from 
a microscopic standpoint, that is, in parameter-free calculations, meaning that the parameters of the 
theory are fixed through the properties of the two- and few-nucleon sytems and never readjusted 
in the many-body system. 
It is the purpose of this paper to present such an investigation. 

Other recent studies of tensor correlations
can be found in Refs.~\cite{Carb13,Rios14}, where 
the self-consistent
Green's function method is used to obtain single-particle properties.       
In Ref.~\cite{Wir+} both single-nucleon and nucleon-pair momentum distributions in 
$A \leq$ 12 nuclei are addressed.                                                      

I also wish to stress that, 
while discussing new aspects and phenomena related to the tensor force, one should not ignore 
what is known since a long time about this important force component. 
First and foremost, its role in the description of NN data and the NN bound state must be taken into account realistically. 
Without such constraint, any discussion on off-shell effects and/or short-range correlations is,
to a large extent, arbitrary. For instance, 
excessive spreading among predictions of the symmetry energy from phenomenological models, such as the numerous versions of the Skyrme model, originates from lack of free-space
constraints and may create an artificially amplified theoretical uncertainty.           

This paper is organized as follows. 
First I review some basic concepts leading to the definition of the defect function and the 
wound integral, both closely related to the correlated wave function. This is done in Sec.~\ref{Descr}.
In Sec.~\ref{dbhf}, 
I proceed with calculations of short-range correlations in nuclear matter within the scheme described in Ref.~\cite{FS14}, which consists of a quantitative meson-theoretic
potential and the Dirac-Brueckner-Hartree-Fock (DBHF) approach to nuclear matter. I pay particular attention           
to tensor and central correlations as seen through the $^3S_1-^3D_1$ channel and the $^1S_0$ state, respectively, and explore their density and isospin-asymmetry
dependence. 

While appreciating the convenience of the DBHF method, I am open to alternative approaches.    
The very popular chiral perturbation theory \cite{Wei68,Wein79} is based on a different philosophy than 
meson theory, and has a firm link to QCD. 
Together with power counting, it allows for a systematic, order-by-order, development of
nuclear forces. Two- and many-body forces emerge naturally, and on an equal footing, at each order of the 
perturbation. Conceptually, this is a very attractive scenario.                                   
I will use a state-of-the-art chiral NN potential along with consistent three-body forces and explore the 
effects of three-body forces on the correlation function. This is accomplished in Sec.~\ref{chi}, where I also 
include 
a discussion on the model dependence of the symmetry energy, extended to a broader range
of interactions. 
A brief summary and conclusions are contained in Sec.~\ref{Concl}.

\section{Short-range correlations: some general aspects} 
\label{Descr} 

In terms of relative and center-of-mass momenta, the Bethe-Goldstone equation can be written as 
\begin{equation}
G({\bf k}_0, {\bf k},{\bf P}^{c.m.}, E_0) = V({\bf k}_0, {\bf k}) +\int d^3{\bf k}^{'} V({\bf k}_0, {\bf k}^{'})        
\frac{Q(k_F,{\bf k}^{'}, {\bf P}^{c.m.})}{E-E_0}           
G({\bf k}^{'}, {\bf k},{\bf P}^{c.m.}, E_0) \; ,                                                                         
\label{BG} 
\end{equation}
where $V$ is the NN potential, $Q$ is the Pauli operator, $E = E({\bf k}^{'}, {\bf P}^{c.m.})$, and 
$E_0 = E({\bf k_0}, {\bf P}^{c.m.})$, with the function $E$ the total energy of the two-nucleon pair. 

The second term of Eq.~(\ref{BG}) represents the infinite ladder sum which builds
short-range correlations (SRC) into the wave function. In the next two equations, I will switch, for simplicity, to operator notation. The correlated ($\psi$) and the uncorrelated ($\phi$) 
wave functions are related through 
\begin{equation}
G\phi = V \psi  \; , 
\label{wf} 
\end{equation}
which implies 
\begin{equation}
\psi = \phi + V \frac{Q}{E-E_0}G\phi \; . 
\label{psi} 
\end{equation}
The difference between the correlated and the uncorrelated wave functions, 
$f=\psi - \phi$, is referred to as the defect function, and is clearly a measure of SRC.                  
It is convenient to consider its momentum-dependent Bessel transform, which gives, for each angular 
momentum state, (and average center-of-mass momentum $P_{avg}^{c.m.}(k_0,k_F)$),             
\begin{equation}
f_{LL'}^{JST}(k,k_0,k_F) = \frac{k \; \bar{Q}(k_F,k,P_{avg}^{c.m.}) G_{LL'}^{JST}(P_{avg}^{c.m.},k,k_0)}{E_0-E} \; , 
\label{ff} 
\end{equation}
where the angle-averaged Pauli operator has been employed. 
This is related to the probability of exciting two nucleons with relative momentum $k_0$ and relative orbital
angular momentum $L$ to a state with 
relative momentum $k$ and relative orbital
angular momentum $L'$.                   
The integral of the probability amplitude squared is known as the wound integral  
and defined, for each partial wave at some density $\rho$, as 
\begin{equation}
\kappa_{LL'}^{JST}(k_0,k_F) = \rho \int_0^{\infty} |f_{LL'}^{JST}(k,k_0,k_F)|^2 dk \; .         
\label{kappa} 
\end{equation}
Thus, $f$ and $\kappa$ provide a clear measure of correlations present in the wave function and the 
$G$-matrix. 

In the present calculations, I take 
the initial momentum equal to 0.55$k_F$. I have chosen it because it is the $r.m.s.$ value of the relative momentum 
of two nucleons having an average center-of-mass momentum, $P_{avg}^{c.m}$, such that their initial
momenta in the nuclear matter rest frame, $k_1$ and $k_2$, are below the Fermi sea. With these constraints,
one can write \cite{HT70} 
\begin{equation}
<k_0^2> = \int _0 ^{k_F} k_0^2 w(k_0, k_F) k_0^2 dk_0 \; ,                                             
\end{equation}
where $w(k,k_F)$ is a weight function whose definition originates from the average 
center-of-mass momentum taken to be \cite{HT70} 
\begin{equation}
(P_{avg}^{c.m.})^2 = \frac{3}{5} k_F^2 \Big ( 1 - \frac{k_0}{k_F} \Big ) \Big (1 + \frac{k_0^2/k_F^2}
{3(2 + k_0/k_F)} \Big ) \;. 
\end{equation}

In isospin-asymmetric matter, it is convenient to work with the total density
$\rho = \rho_n + \rho_p$ and the asymmetry (or neutron excess) parameter
$\alpha = \frac{ \rho_n - \rho_p}{\rho}$, where 
$\alpha$=0 corresponds to symmetric matter and 
$\alpha$=1 to neutron matter.                       
In terms of $\alpha$ and the average Fermi momentum, $k_F$, related to the total density in the usual way, 
namely, 
\begin{equation}
  \rho =\frac{2 k_F^3}{3 \pi ^2} \; ,   \label{rho}   
\end{equation}
the neutron and proton Fermi momenta can be expressed as 
\begin{equation}
 k^{n}_{F} = k_F{(1 + \alpha)}^{1/3} \;\;\;\; \mbox{and} \; \;\; \;                      
 k^{p}_{F} = k_F{(1 - \alpha)}^{1/3} ,            \label{kfp} 
\end{equation}
 respectively.

The $G$-matrices in this work are isospin sensitive in the sense the $G_{nn}$, $G_{pp}$, and 
$G_{np}$ are different even in the same isospin state, due to the different Fermi 
momenta of neutrons and protons.                                                            
In the end, the self-consistent procedure \cite{FS14} provides 
the single-neutron and single-proton potentials together with the 
$G$-matrices $G_{ij}$, $ij=nn$, $pp$, or $np$. 

In short, the defect function Eq.~(\ref{ff}) can be calculated for a particular pair of nucleons, ($np$, $pp$, or $nn$), using the appropriate $G$-matrix and Pauli operator. The latter depends only on the neutron(proton) Fermi 
momentum, $k_F^n$$(k_F^p)$, in the case of $nn$($pp$) scattering; or, it is the asymmetric Pauli operator
for two particles with different Fermi momenta in the $np$ case, $\bar{Q}(k_F^n,k_F^p)$.                                

\section{The Dirac-Brueckner-Hartree-Fock approach} 
\label{dbhf}  
\subsection{Brief review} 
\label{tech}  
A popular approach to nuclear matter is constructed from non-relativistic NN potentials complemented 
by three-body forces. An example of this method can be found in Ref.~\cite{Catania}. 
Local potentials together with phenomenological three-body forces \cite{av18} have also been widely used. 
I will come back to those later in the paper (Sec.~\ref{esymm}). 
As an alternative, 
relativistic approaches to nuclear matter have been pursued through the Dirac-Brueckner-Hartree-Fock
(DBHF) scheme \cite{FS14}. The main strength of this framework is in its inherent ability to account for 
an important class of three-body forces (3BF) which turns out to be crucial for nuclear matter saturation, namely, 3BF 
arising from the virtual excitation of nucleon-antinucleon intermediate pairs.           
My standard choice for the NN interaction is the Bonn B potential \cite{Mac89}, a relativistic 
potential which uses pseudovector coupling for the coupling of pseudoscalar mesons with nucleons.
Details of the DBHF approach as I apply it to asymmetric matter can be found in Ref.~\cite{FS14}.

\subsection{Predictions with Dirac-Brueckner-Hartree-Fock} 
\label{pred}  

\begin{figure}[!t] 
\centering         
\vspace*{-1.2cm}
\hspace*{0.5cm}
\scalebox{0.35}{\includegraphics{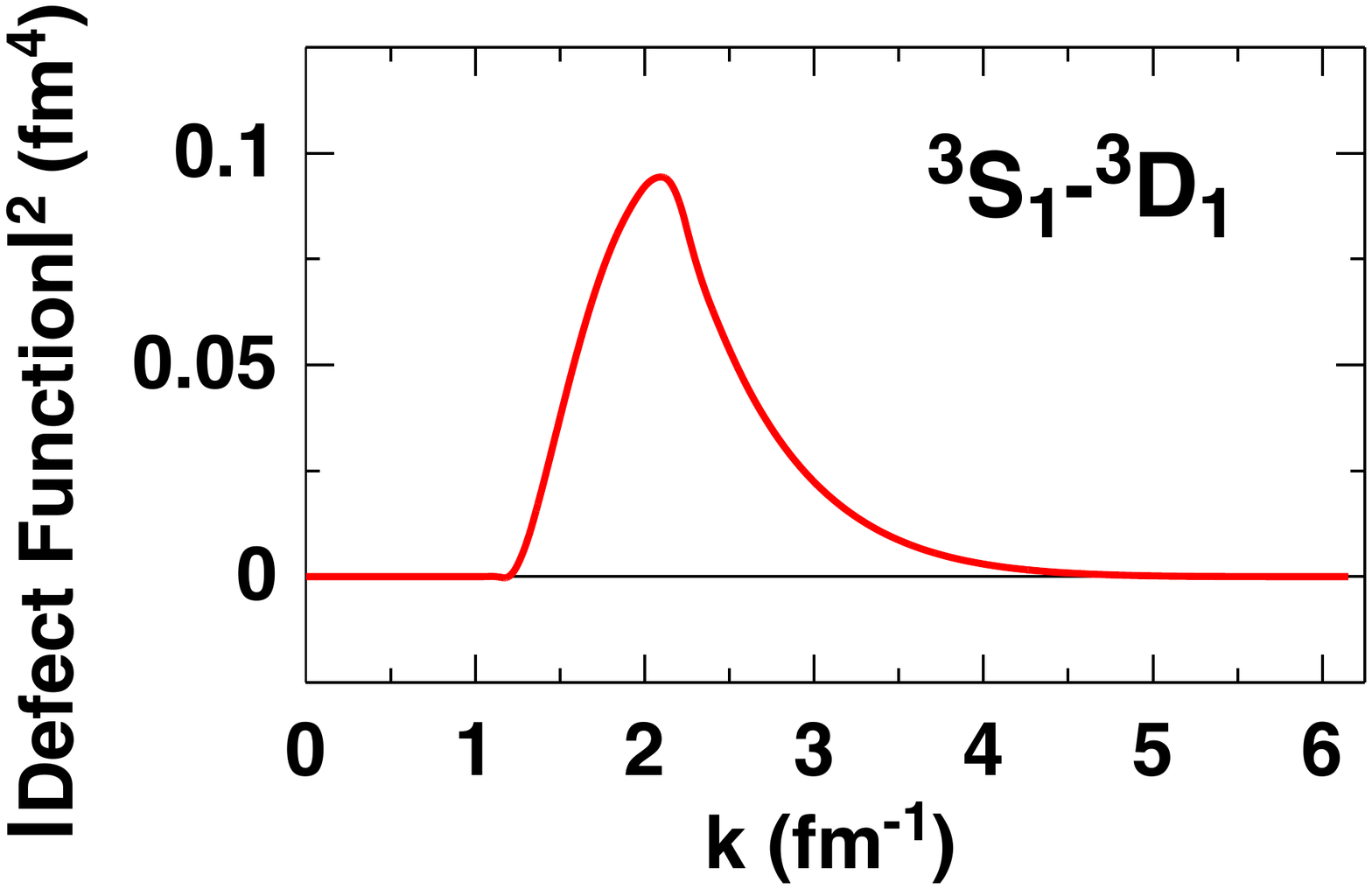}}
\scalebox{0.35}{\includegraphics{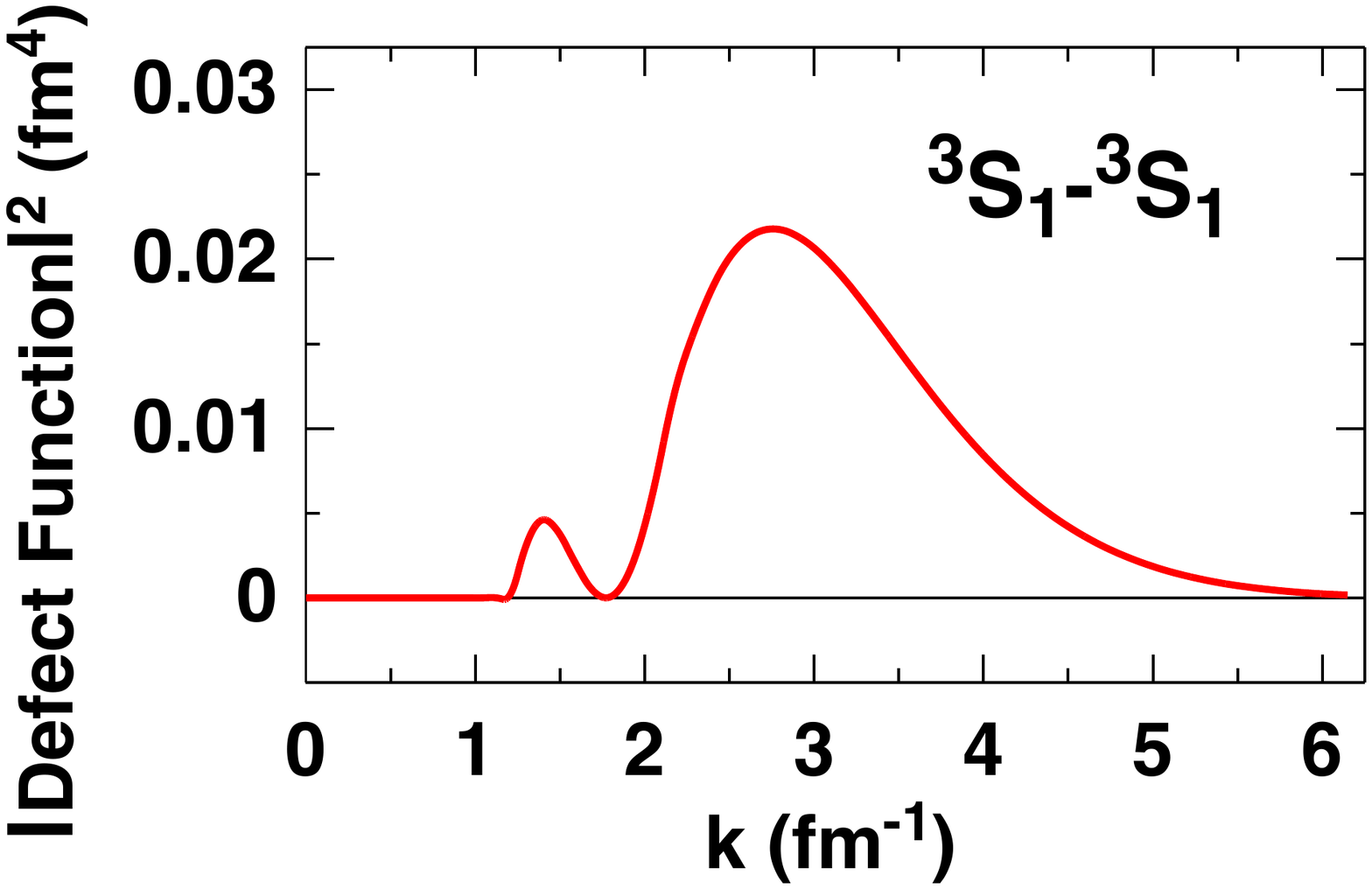}}
\vspace*{-3.9cm}
\caption{(Color online)                                                         
Magnitude squared of the defect function Eq.~(\ref{ff}) for the $^3S_1-^3D_1$ transition (left) and the $^3S_1$ state (right) in symmetric nuclear matter. The Fermi momentum                 
is equal to 1.4 fm$^{-1}$.                                                                      
}
\label{sd14}
\end{figure}

I begin by showing on the left-hand side of Fig.~\ref{sd14} the magnitude squared of the 
defect function, Eq.~(\ref{ff}), for the $^3S_1-^3D_1$ transition                                                      
as a functions of the final relative momentum $k$ in symmetric matter.
The total density corresponds to 
a Fermi momentum of 1.4 fm$^{-1}$.                                                                    
Notice that these distributions are 
excitation probabilities rather than standard momentum distributions (which are usually larger 
at low momenta). In other words, these curves do not include the distribution of momenta for 
occupied states below the Fermi surface. 

Clearly there is 
a high probability that the $np$ pair is excited to a state with relative momentum 
of about 2 fm$^{-1}$ {\it via} a tensor transition. 
For comparison, I show on the right-hand side of Fig.~\ref{sd14} the same quantity for 
the $^3S_1$ state.                                      
Note that the latter carries information on short-range central correlations, namely the repulsive core of the central force, 
although it is also impacted by the tensor force because of its 
coupling to the $D$-state. 
It peaks around a momentum of about 
3 fm$^{-1}$ and has a distinct node between                                              
1.5 and 2 fm$^{-1}$.                                                                    
Also, the $^3S_1$ probability amplitude tends to be broader, that is, it ``survives" higher momenta, or 
shorter ranges. 
Notice that the amplitude of the tensor transition has a much larger absolute value.                                   

\begin{figure}[!t] 
\centering         
\vspace*{1.2cm}
\hspace*{0.5cm}
\scalebox{0.5}{\includegraphics{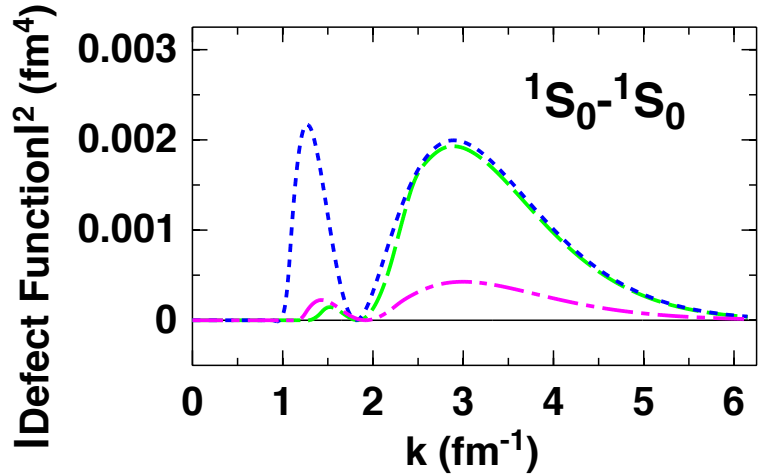}}
\vspace*{-5.5cm}
\caption{(Color online)                                                                         
Magnitude squared of the defect function Eq.~(\ref{ff}) for the $^1S_0$ state in asymmetric matter with 
$\alpha$=0.4. The Fermi momentum corresponding to the total nucleonic density 
is equal to 1.4 fm$^{-1}$. The dotted (blue), dashed (green), and dash-dotted (purple) curves are calculated
using the $pp$, $nn$, and $pn$ $G$-matrices, respectively, with the appropriate Pauli operators and isospin coefficients. See text 
for more details.                   
} 
\label{1s0asy}
\end{figure}

In Fig.~\ref{1s0asy}, I focus on correlations in the 
$^1S_0$ partial wave, which is accessible to $pn$ as well as $nn$ and $pp$ pairs. Obviously, in this state SRC originate
from the repulsive core of the central force. The dotted (blue), dashed (green), and dash-dotted (purple)
curves are the magnitude squared of Eq.~\ref{ff} obtained with the $pp$, $nn$, and $np$ $G$-matrices, 
respectively, as outlined in Section~\ref{Descr}. This is asymmetric matter, with neutron excess
parameter equal to $0.4$, and the Fermi momentum (corresponding to the {\it total nucleon density}) equal to
1.4 fm$^{-1}$. The neutron and proton Fermi momenta are then given by Eq.~(\ref{kfp}). 

The smaller size of the $np$ probability amplitude as compared to the one in the $nn$ or $pp$ cases 
is not in contraddiction with the considerations made in the Introduction, nor should it be surprising.
While a statistical factor of $2T+1$ is present in the isospin-saturated system, 
the three cases in Fig.~\ref{1s0asy} have {\it definite} isospin coefficients. Specifically, the appropriate factor  
by which the $G$-matrix must be multiplied for the 
coupling of the two nucleon isospins in a state of $T=1$ and $T_z=\pm 1$ is equal to 1, whereas 
for two nucleons coupling to a state of $T=1$ and $T_z=0$ it is equal to $(\frac{1}{\sqrt{2}})^2$.
Therefore, the $np$ curve in Fig.~\ref{1s0asy}, being proportional to the $G$-matrix squared, should be at least a factor of four smaller than the other
two. 
The much larger 
degree of correlation seen in experiments among $np$ pairs is due to interactions in 
$^3S_1-^3D_1$ (see Fig.~\ref{sd14}), not $^1S_0$.                              

Before moving to the next figure, one should take note of the enhancement of the $pp$ defect function at the lower momenta,
due to the proton lower Fermi momentum in neutron-rich matter            
(see Eqs.~(\ref{kfp}), and thus weaker Pauli blocking.

\begin{figure}[!t] 
\centering         
\vspace*{1.2cm}
\hspace*{0.5cm}
\scalebox{0.25}{\includegraphics{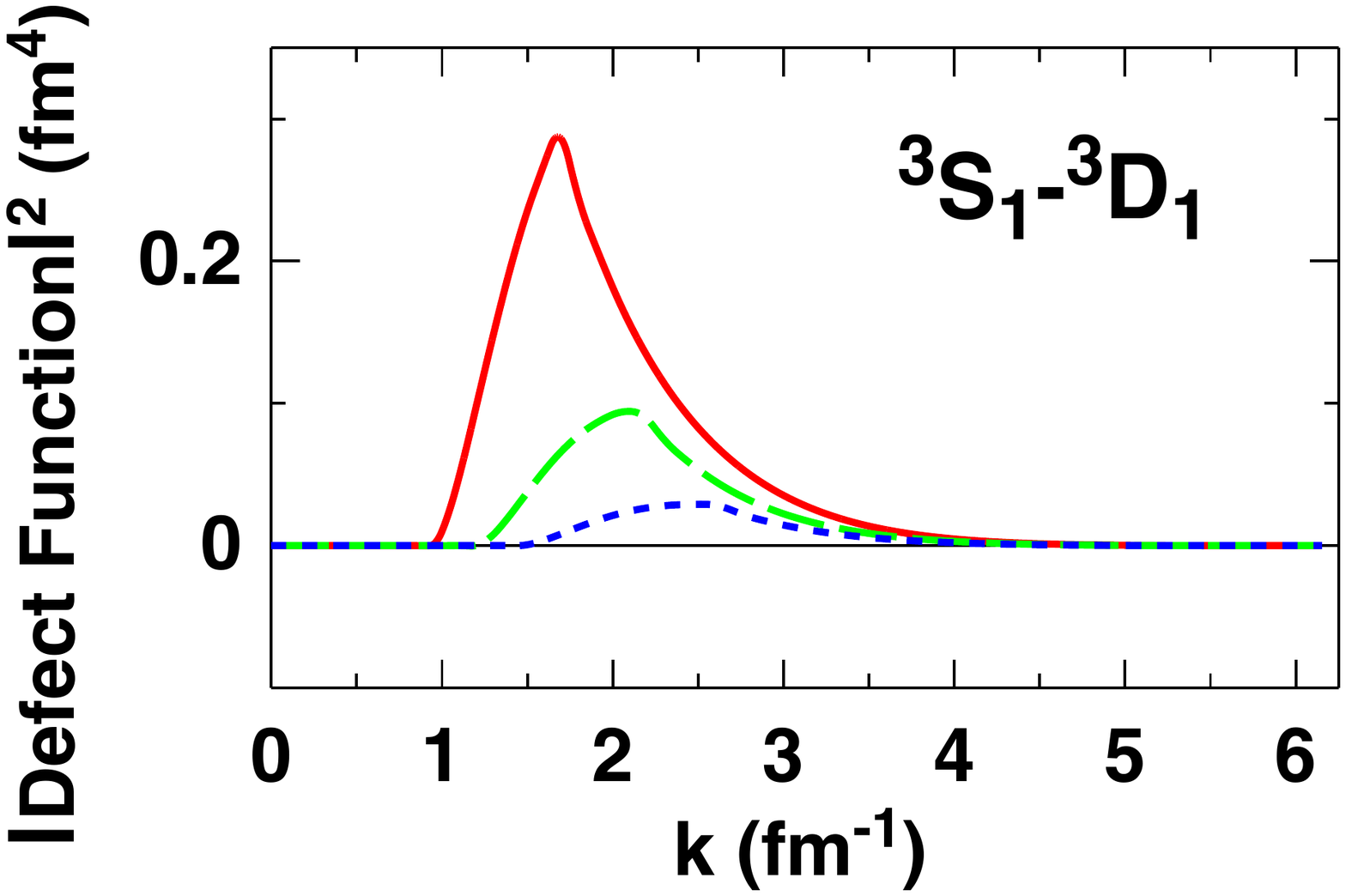}}
\scalebox{0.25}{\includegraphics{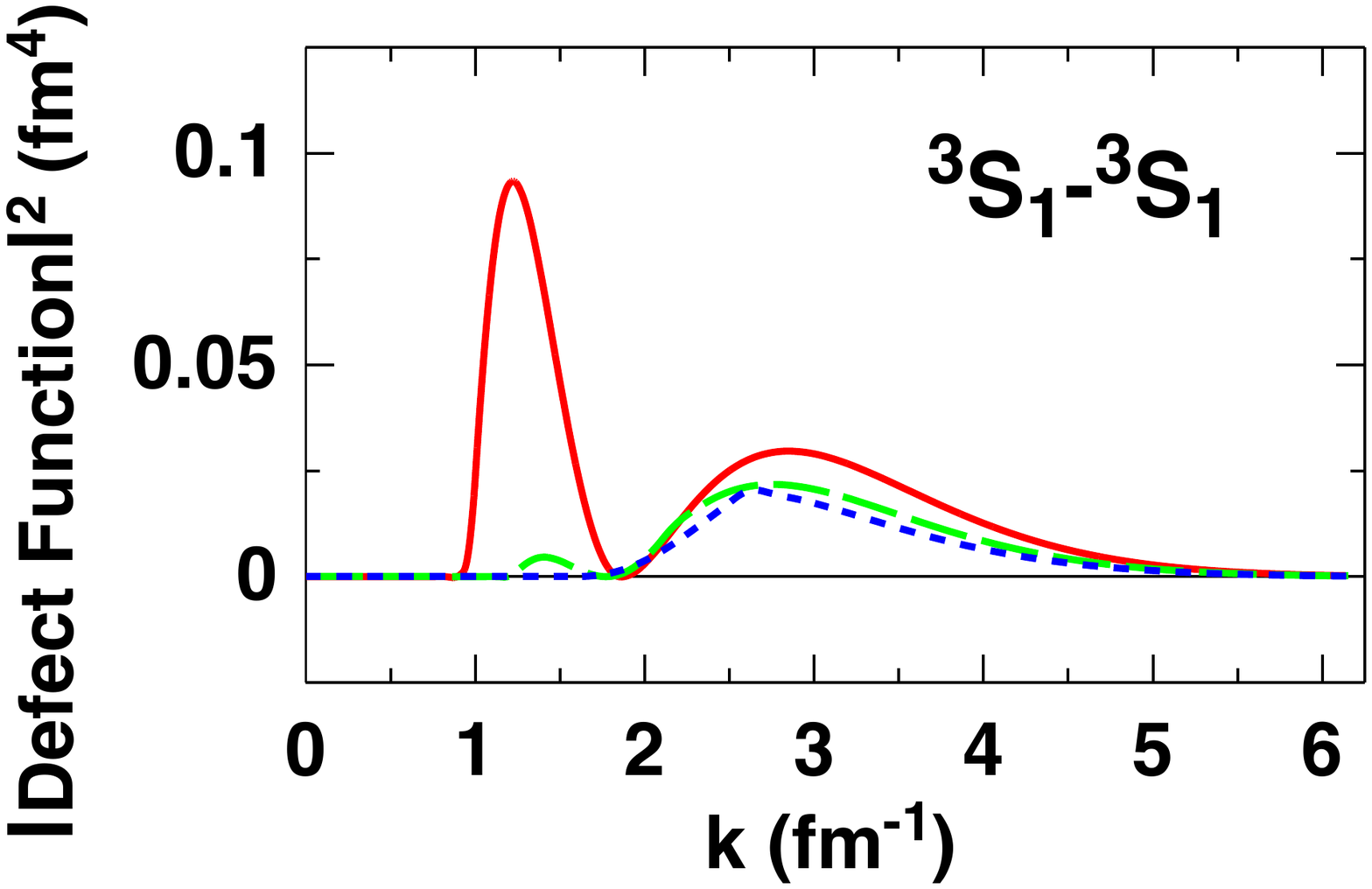}}
\scalebox{0.25}{\includegraphics{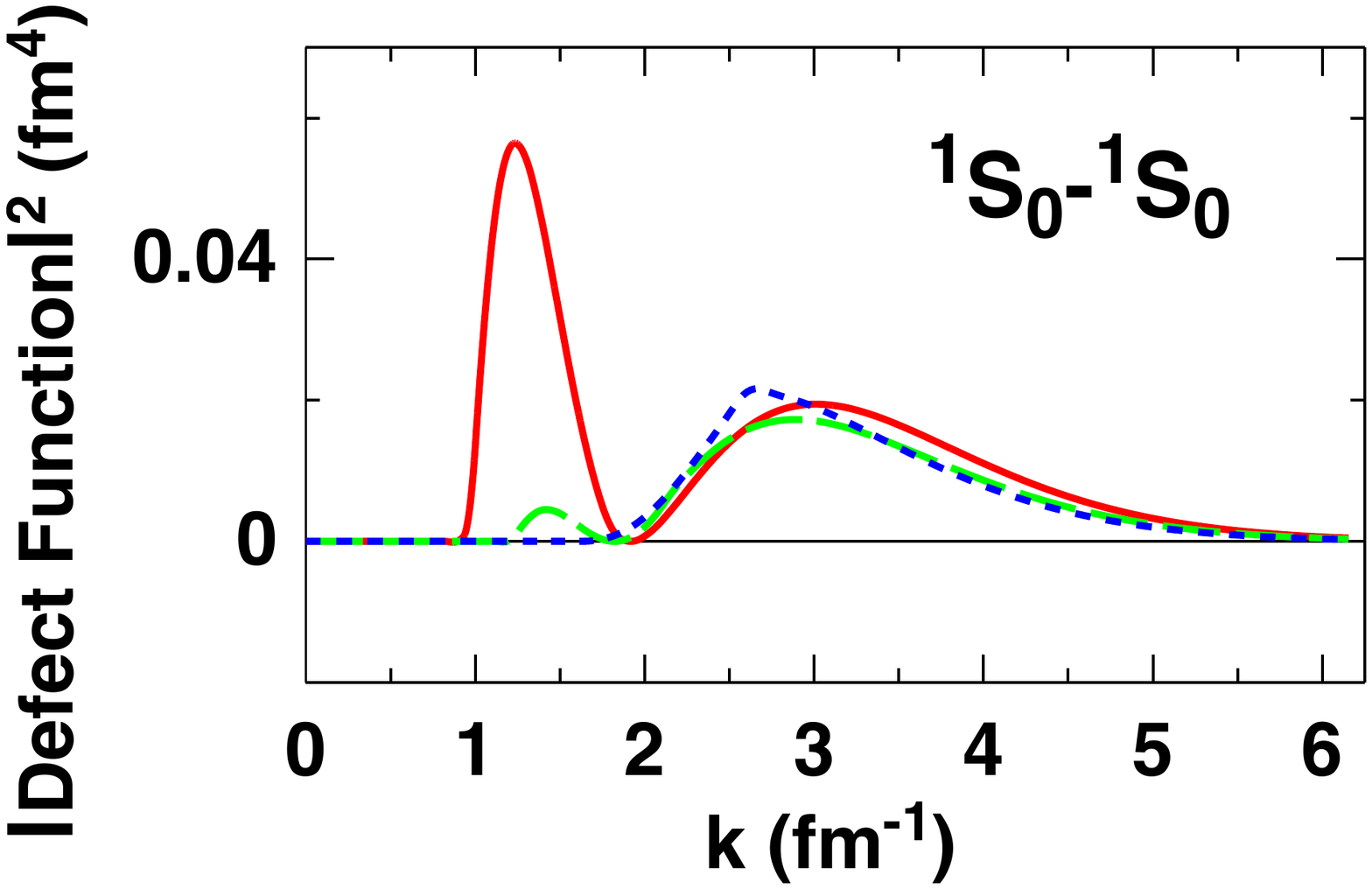}}
\vspace*{-1.8cm}
\caption{(Color online)                                              
The magnitude squared of Eq.~\ref{ff} for three different values of the Fermi
momentum in symmetric matter: $k_F$=1.1 fm$^{-1}$ (solid red); 
$k_F$=1.4 fm$^{-1}$ (dashed green); 
$k_F$=1.7 fm$^{-1}$ (dotted blue). 
}
\label{all}
\end{figure}

In Fig.~\ref{all} I consider three different densities of symmetric nuclear matter.                    
The defect function for the tensor transition
maintains a similar shape with changing density, with the peak shifting towards lower(higher) momenta at the        
lower(higher) density, due to the changing impact of Pauli blocking in each case.                        
 For both $^3S_1$ and $^1S_0$, the peak at the lower momenta grows large at the lower density.    

The individual contributions to the wound integral, Eq.~\ref{kappa}, from the states considered in the figures are shown 
in Table~\ref{tab1} for three densities of symmetric matter. The contribution of the central force (as seen
through $^1S_0$) relative to the tensor force increases with increasing density, due to the enhanced
impact of the repulsive core when higher momenta are probed (as is the case in a system with increasing 
Fermi momentum). 

\begin{table}                
\centering \caption                                                    
{Contributions to the wound integral, Eq.~\ref{kappa}, from $J=0$ and $J=1$ states at different densities. 
} 
\vspace{5mm}
\begin{tabular}{|c|c|c|c|}
\hline
$ k_F$ (fm$^{-1}$) & $^3S_1-^3D_1$ & $^3S_1-^3S_1$ & $^1S_0$ \\
\hline     
 1.1 &  0.079& 0.025 & 0.017   \\
\hline
 1.4 &  0.060 & 0.022  & 0.019 \\ 
\hline
 1.7 &  0.037 & 0.031 & 0.035 \\ 
\hline
\end{tabular}
\label{tab1} 
\end{table}

\section{Chiral Interactions}
\label{chi}
\subsection{Some general aspects}

Ideally, one wishes to base a derivation of the nuclear force on QCD. However, the well-known 
problem with QCD is that it is non-perturbative in the low-energy regime characteristic for 
nuclear physics. For many years this fact was perceived as a great obstacle for a derivation of
nuclear forces from QCD, impossible to overcome except with lattice QCD. The effective field theory
concept has proposed a way out of this dilemma. One has to realize that the scenario of low-energy QCD
is characterized by pions and nucleons interacting via a force governed by spontaneously broken
approximate chiral symmetry \cite{Wei68,Wein79}. For a recent review, the reader is referred to Ref.~\cite{ME11}. 

Before proceeding, it is appropriate to point out some of the extensive literature concerned with chiral
dynamics in isospin symmetric and asymmetric nuclear matter. 
The work by the Munich group (see, for instance, Refs.~\cite{Kais02,Kais12}), was recently reviewed in Ref.~\cite{Mun13}.
Other authors have adopted chiral low-momentum interactions to soften the short-range components
of the original potentials. A survey of renormalization group methods and their connection to chiral
effective theory, as well as applications to nuclear matter, can be found in Refs.~\cite{Bogn10,Hebe11}.
In Ref.~\cite{Barc13}, the authors report symmetric nuclear matter predictions obtained with chiral interactions
within the self-consistent Green's function approach. Three-body forces are included through effective 
one-body and two-body interactions computed from averaging over the third nucleon. 

Here, I will use a high-precision NN potential at next-to-next-to-next-to-leading order \cite{EM03}. 
For chiral interactions                  
the characteristic momentum scale is below the scale set by the cutoff in the 
regulator function. For the interaction employed here, 
the latter has the form 
\begin{equation}
f(p',p) = exp[-(p'/\Lambda)^{2n} - (p/\Lambda)^{2n}] \; .
\label{reg}
\end{equation}

The low-energy constants $c_{1,3,4}$ 
associated with the $\pi \pi N N$ contact couplings of the ${\cal L}^{(2)}_{\pi N}$ chiral
Lagrangian can be fitted to $\pi N$ or $NN$ scattering data. Their values are given in Table~\ref{tabci}.         

\begin{figure}[!t] 
\centering         
\vspace*{1.5cm}
\hspace*{-0.7cm}
\scalebox{1.0}{\includegraphics{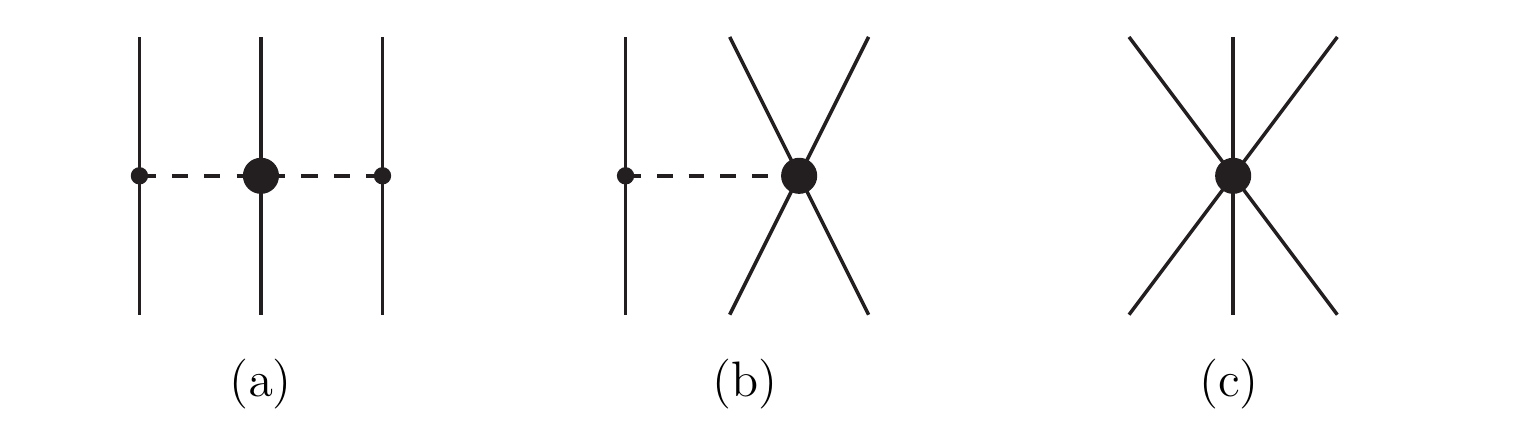}}
\vspace*{0.5cm}
\caption{(Color online) Leading three-body forces at N$^2$LO. See text for more details.     
} 
\label{3b}
\end{figure}

The three-nucleon forces which make their appearance at the third order in the chiral power 
counting (N$^2$LO), are: the long-range two-pion exchange graph;
the medium-range one-pion exchange diagram; the short-range contact. The corresponding diagrams are 
shown in Fig.~\ref{3b}(a), \ref{3b}(b), and \ref{3b}(c), respectively. 
A total of six one-loop diagrams contribute at this order. Three are generated by the two-pion 
exchange graph of the chiral three-nucleon interaction and depend on the low-energy constants
$c_{1,3,4}$, which are fixed in the NN system \cite{EM03} as explained above.                 
Two are generated by the one-pion exchange diagram and depend on the low-energy constant 
$c_D$. Finally, the short-range component depends on the constant $c_E$. 
In pure neutron matter, the contributions proportional to the low-energy constants $c_4,c_D$, and 
$c_E$ vanish \cite{Holt}. 

Although order consistency would require both two-body forces (2BF) and 3BF at 
N$^3$LO, such calculation for nuclear matter is not feasible at this time. 
Thus the combination of 2BF at 
N$^3$LO and 3BF at                                       
N$^2$LO is presently state-of-the-art. It should be noted that four-body forces also appear at this order \cite{4BF1}, but are left 
out because they are expected to be small \cite{4BF2}.

Concerning the 
low-energy constants 
$c_D$ and $c_E$ appearing in the 
N$^2$LO 3BF,                                                                                  
a very important aspect of these calculations is that       
they are completely determined from the three-nucleon system. Specifically,            
they are constrained to reproduce the A=3 binding energies and the triton Gamow-Teller matrix elements.
The procedure \cite{Marc} is based on consistency of 2BF, 3BF, and currents, as required by chiral EFT. 
Their values are given in Table~\ref{tabci}.

\begin{table}
\centering
\begin{tabular}{|c||c|c|c|c|c|c|c|}
\hline
Order & $\Lambda$ (MeV) & $n$& $c_1$ & $c_3$ & $c_4$ & $c_D$ & $c_E$ \\ 
\hline     
N$^3$LO & 450 & 3& -0.81 & -3.40 & 3.40 &-0.24 &-0.11  \\
\hline
\hline
\end{tabular}
\caption{Values of $n$ and $\Lambda$                                                                    
used in the regulator function, Eq.~(\ref{reg}),      
low-energy constants of the dimension-two $\pi N$ Lagrangian   
$c_{1,3,4}$ (given in                                 
units of GeV$^{-1}$), and 
low-energy constants $c_D$ and $c_E$ as used in the three-body force.} 
\label{tabci}
\end{table}

In Ref.~\cite{Holt}, density-dependent corrections to the in-medium NN interaction have been derived from              
the leading-order chiral 3BF. These are 
effective two-nucleon interactions obtained from the underlying three-nucleon forces by
integrating one nucleon up to the Fermi momentum. 
Therefore, they are 
computationally very convenient.                                                            
Analytical expressions for these corrections are provided in Ref.~\cite{Holt} in terms of the well-known 
non-relativistic two-body nuclear force operators, which  
can be conveniently 
incorporated in the usual NN partial wave formalism and the conventional BHF theory.

\subsection{Predictions with chiral interactions}
\label{chipred} 
\begin{figure}[!t] 
\centering         
\vspace*{1.2cm}
\hspace*{0.5cm}
\scalebox{0.25}{\includegraphics{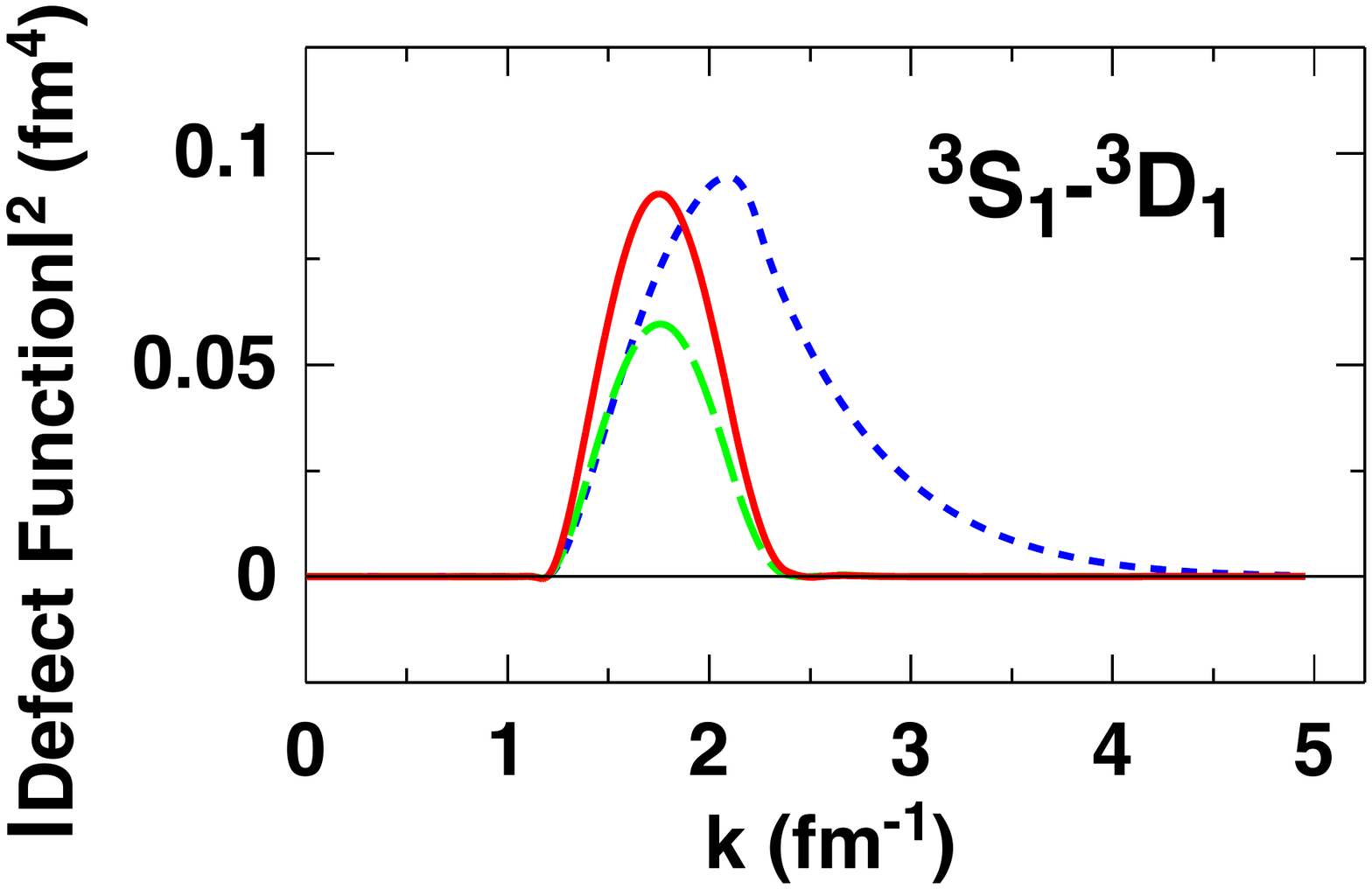}}
\scalebox{0.25}{\includegraphics{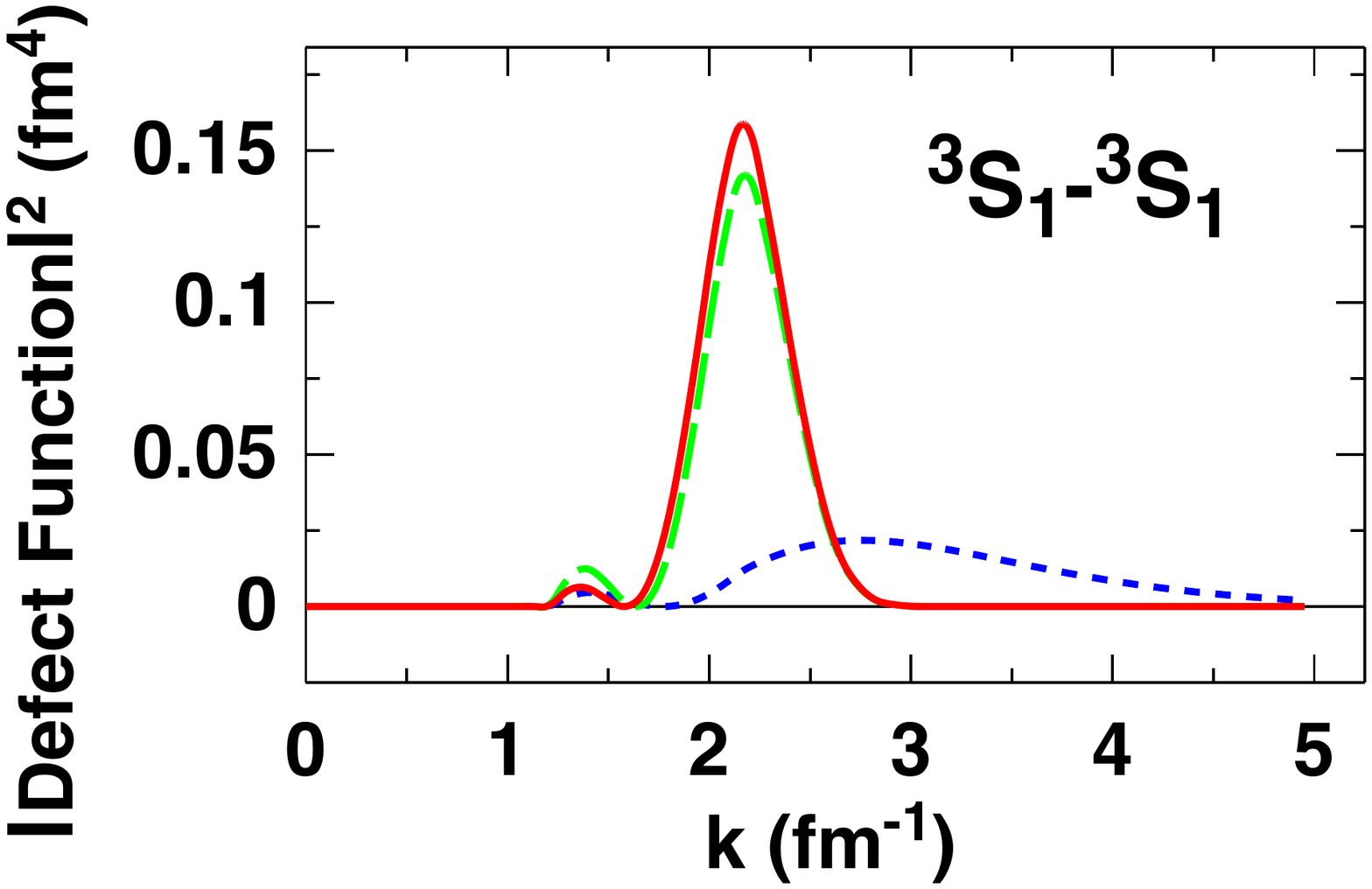}}
\scalebox{0.25}{\includegraphics{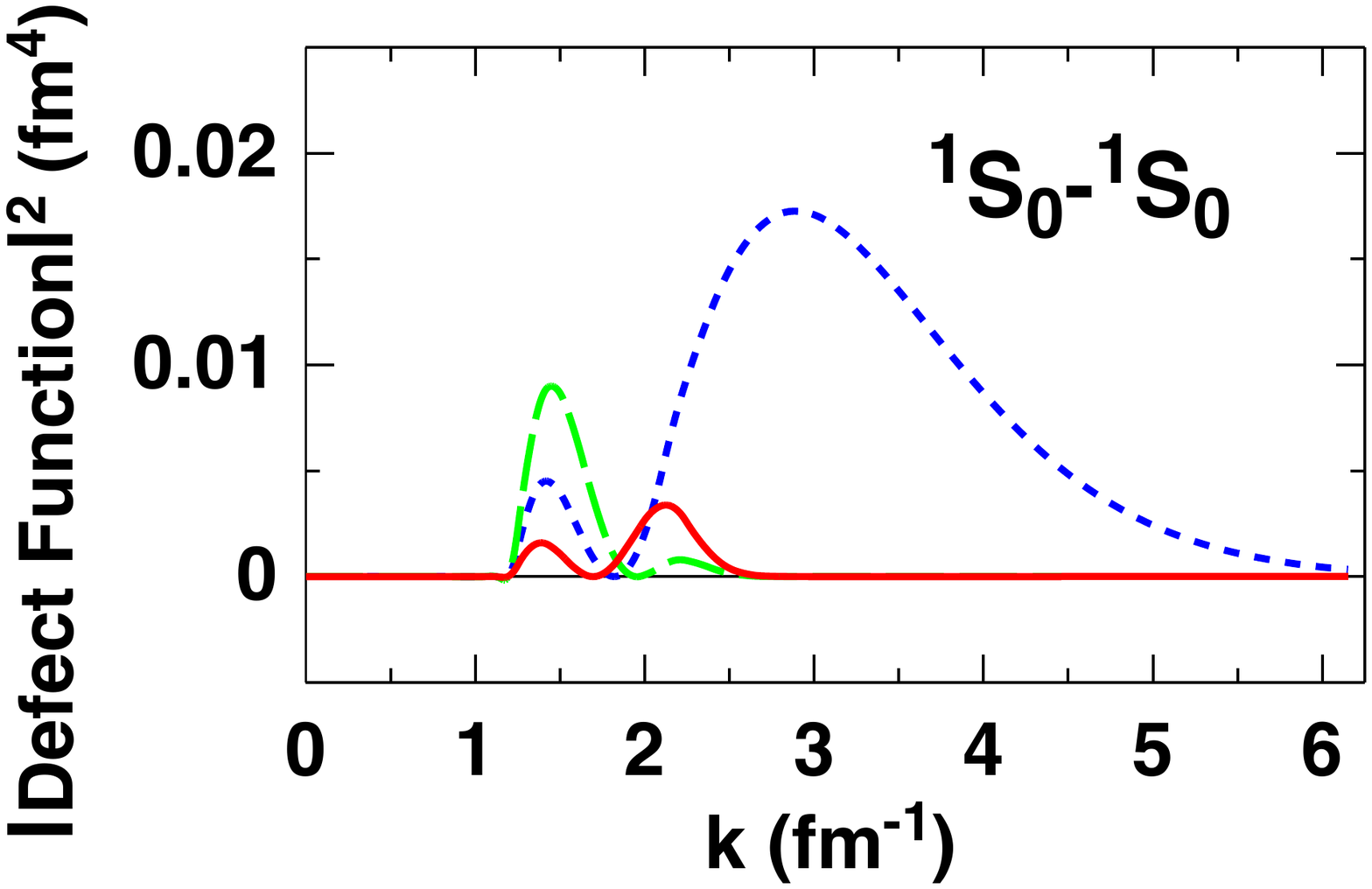}}
\vspace*{-1.8cm}
\caption{(Color online)                                              
Blue (dotted): standard prediction of the magnitude squared of the defect function 
from the DBHF calculations together with the Bonn B potential; green (dashed): prediction 
with the chiral two-body interaction only; red (solid): prediction with two- and 
three-body chiral interactions. 
Symmetric nuclear matter with Fermi momentum equal to 1.4 fm$^{-1}$. 
}
\label{2b3b}
\end{figure}

In Fig.~\ref{2b3b}, the blue (dotted) curve shows the predictions 
from Bonn B already discussed in the previous Section; the green (dashed) curve displays the prediction 
with the chiral two-body interaction only; finally, the red (solid) curve is obtained with two- and 
three-body chiral interactions. 
In all cases, the Fermi momentum is equal to 1.4 fm$^{-1}$. 

\begin{table}                
\centering \caption                                                    
{The wound integral $\kappa$ in symmetric nuclear matter (SNM) and neutron matter (NM) for the three calculations shown in Fig.~\ref{2b3b}. 
The total density is the same in both SNM and NM, and equal to 0.185 fm$^{-3}$.
} 
\vspace{5mm}
\begin{tabular}{|c|c|c|}
\hline
Theoretical approach &   SNM  & NM    \\ 
\hline
  Bonn B + DBHF &  0.130   & 0.0133  \\                                       
\hline     
Chiral NN (2BF) & 0.075 & 0.0011 \\
\hline
Chiral NN + 3BF & 0.099 & 0.003  \\
\hline
\hline
\end{tabular}
\label{tab2} 
\end{table}

Short-range correlations with chiral or meson-theoretic interactions can be dramatically
different. In particular, chiral potentials tend to produce a much more localized ditribution of
momenta. This is reasonable since, due to the applied cutoff, chiral potentials are much softer 
that meson-theoretic ones.                                                                 
More precisely, the Bonn B potential vanishes (regardless the partial wave) around 
2000 MeV (in terms of the relative momentum), 
whereas the chiral NN interaction is essentially negligible already near 800 MeV. 
Accordingly, the curves obtained with Bonn B extend to much higher momenta.         

Clearly, the chiral 3BF contributes to the tensor force. For instance, 
for the $^3S_1-^3D_1$ transition near normal density, it increases the probability amplitude 
around 1.5-2 fm$^{-1}$ by about 30\%. 

Interestingly, for the $^3S_1$ state, which contains contributions from the                                      
 the $^3S_1-^3D_1$ intermediate state,                                                      
 both curves obtained with chiral interactions are much larger 
than the Bonn B predictions, while the opposite is true for the 
$^1S_0$ state. Concerning the latter, strictly speaking a ``hard core" (that is, short-range repulsion from the central force),
can only be defined for a local potential. Chiral potentials have a higher degree of non-locality as compared
to meson-theoretic potentials (even non-local ones, such as the relativistic Bonn B). This is mostly due to the 
form factor applied to chiral potentials (typically, Gaussian functions of $p$ and $p'$), whereas the form factor
used with meson-theoretic potentials is a function of the momentum transfer and is, therefore, local. 
This may be the reason for the very different structure of the probability amplitude seen in      
central force dominated $S$ waves.                                                                                                         
Next, I will make a comparison between symmetric matter and pure neutron matter through the wound integral including all states. When people calculate single-nucleon or nucleon-pair momentum distributions, typically a much stronger depletion of states 
below the Fermi surface is observed in symmetric matter as compared to neutron matter \cite{Carb13}, indicating the absence        
of short-range tensor correlations from the latter. 
Recalling that $\kappa$ measures the probability of unoccupied states below the Fermi surface, with the present method one   
can gain access to similar information. 
In Table~\ref{tab2} I compare the wound integral in symmetric matter and pure neutron matter around normal density for the three approaches
considered in Fig.~\ref{2b3b}. 
First, one can see that $\kappa$ in symmetric matter is between one and two orders of magnitude larger than in       
neutron matter. 
This is reasonable. First, by far the main contribution to $\kappa$ in neutron matter comes 
from $^1S_0$, which is about three times smaller than the equivalent contribution in symmetric matter due to the
statistical factor of $2T+1$ appearing in the latter case.                                                          
Most important, of the total 
strength of $\kappa$ in symmetric matter, I have observed that almost all of it (nearly 12\% out of 13\%), comes from $J=1$ states, particularly 
 $^3S_1-^3D_1$, absent from neutron matter.                                             
Also, the degree to which SRC are stronger in symmetric matter as compared to neutron matter is very model dependent.

\subsection{The symmetry energy}
\label{esymm} 
To conclude, and to reconnect with some of the statements made in the Introduction, in this section I address the role of the tensor force on the symmetry energy. 
Although the previous sections have focussed on the comparison between a representative relativistic, 
meson-theoretic potential and a state-of-the-art chiral potential with chiral 3BF, here I will expand
the scopes of this study and consider other approaches as well.

It may be useful to recall that, 
in the parabolic approximation, the equation of state of isospin asymmetric matter is 
\begin{equation}
e(\rho,\alpha) = e_0(\rho) + e_{sym}(\rho)\alpha^2 \; . 
\label{eee} 
\end{equation}
Thus the symmetry energy is approximately the difference beween the energy per particle in neutron matter and
in symmetric matter. 

As mentioned earlier, the combination of a local potential and 3BF such as 
the Urbana IX force is broadly used in equation of state calculations and other applications.
Concerning chiral interactions, 
NN potentials with different cutoffs in the regulator function, Eq.~(\ref{reg}), 
can differ considerably in their off-shell behavior. Thus, I will consider two additional chiral 
potentials with different cutoff parameters, specifically $\Lambda$=500 and 600 MeV \cite{EM03}, with the respective
low-energy constants given in Table~\ref{tabnew}. It's important to stress that, for each cutoff, the fit to the 
NN data and the properties of the A=3 system is regained \cite{Marc}. 

\begin{figure}[!t] 
\centering         
\vspace*{1.2cm}
\hspace*{0.5cm}
\scalebox{0.5}{\includegraphics{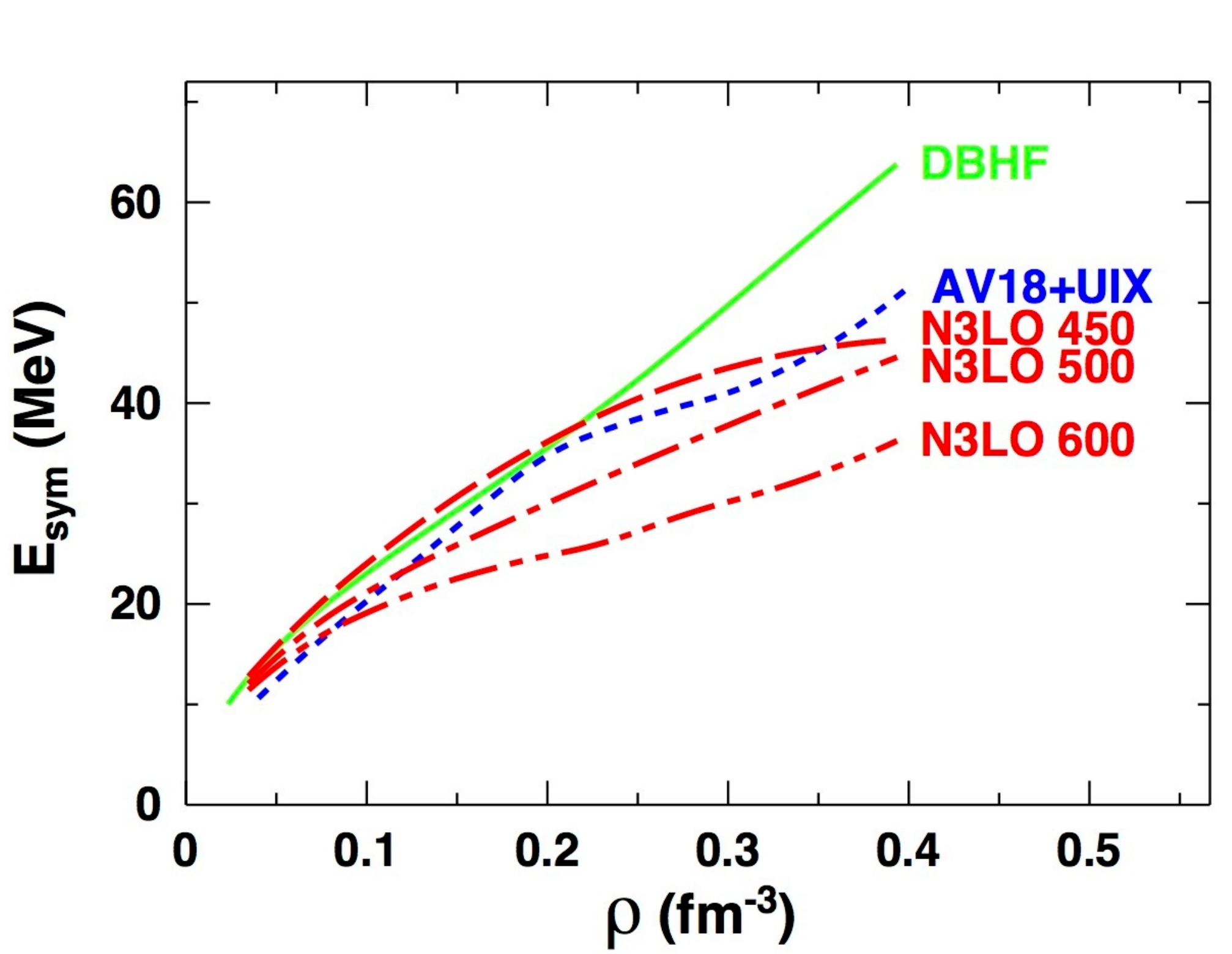}}
\vspace*{0.5cm}
\caption{(Color online)                                                                         
The symmetry energy as a function of density predicted with various interaction models as  
explained in the text. 
} 
\label{esym}
\end{figure}

Figure~\ref{esym} displays the symmetry energy from the following models:     
my standard DBHF predictions (solid green); the Argonne V18 potential with the Urbana IX 3BF \cite{av18} (dotted blue);
three chiral potentials with different, increasing, values of the cutoff (dashed, dash-dotted, and 
dash-double-dotted). 
Thus, the figure represents a broad spectrum of interactions, ranging from (most repulsive) 
relativistic meson theory to the (softer) chiral forces with 
different ``UV scales".        
Notice that the chiral potentials with larger cutoff appear softer as seen through the symmetry 
energy because the equation of state of {\it symmetric} nuclear matter becomes considerably more repulsive
with increasing cutoff (that is, at a faster rate than neutron matter). 
Recalling Eq.~(\ref{eee}), and the fact the isospin-zero states are absent from neutron matter, 
this points again to the chief role of the tensor force.

In summary, 
keeping in mind that different interactions differ mostly in their high-momentum components, 
Fig.~\ref{esym} demonstrates in a remarkably clear way how the high-momentum differences, such as those which have been explored in the previous sections, 
impact the high-density behavior of the symmetry energy.

\begin{table}
\centering
\begin{tabular}{|c||c|c|c|c|c|c|c|}
\hline
\hline
N$^3$LO & $\Lambda$ (MeV) & $n$& $c_1$ & $c_3$ & $c_4$& $c_D$ & $c_E$ \\
\hline     
 & 500 & 2& -0.81 & -3.20 & 5.40 & 0.0 & -0.18 \\
 & 600 & 2& -0.81 & -3.20 & 5.40  & -0.19 & -0.833 \\
\hline
\hline
\end{tabular}
\caption{Values of $n$ and low-energy constants for different values of the cutoff.                   
The 
$c_{1,3,4}$ LECs                                                                            
are given in units of GeV$^{-1}$.}
\label{tabnew}
\end{table}

\section{Summary and Conclusions}                                                                  
\label{Concl} 

The magnitude squared of the so-called defect function is 
closely related to the nucleon-pair momentum distribution and is a measure of short-range correlations. 
I have examined such correlations through 
the coupled and uncoupled $S$-waves, which are mostly
impacted by the short-range tensor and/or central forces. 

I have focussed on two different microscopic approaches. In one case, a well-known 
meson-theoretic NN potential is employed in a relativistic calculation of the nuclear matter 
$G$-matrix. In the other, modern two- and three-body chiral forces are used in calculations of
the Brueckner $G$-matrix. Notice that the theoretical basis of the input nuclear forces is fundamentally 
different in each of the two {\it ab initio} approaches. 

Short-range correlations 
depend strongly on the nature of the underlying nuclear forces. The momentum distributions with 
meson-theoretic or chiral forces (whether 3BF are included or not) are different both quantitatively and qualitatively, with 
chiral interactions yielding characteristically more localized distributions. 

Chiral three-body forces have a large impact on SRC.                                                               
In the $^3S_1-^3D_1$ transition at normal density, they increase the probability amplitude 
around 1.5-2 fm$^{-1}$ by about 30\%. 
In the $^1S_0$ state, chiral interactions appear much softer than the meson-theoretic one, possibly due to 
a high degree of non locality in the chiral interactions.                 

Concerning isospin dependence, 
short-range correlations                  
are negligible in neutron matter as compared to symmetric matter. 

Last, I observed that the high-density behavior of the symmetry energy                                           
is impacted strongly by the nature of the nuclear force at high momentum. 

I would like to conclude by highlighting once again the importance of 
microscopic approaches                                                                       
as opposed to phenomenological ones. The former 
give insight into the role 
of specific components of
nuclear forces and their behavior in the medium, with the reassuring 
awareness that these forces are constrained by free-space NN data and the properties of 
few-body systems.

\section*{Acknowledgments}
This work was supported by 
the U.S. Department of Energy, Office of Science, Office of Basic Energy Sciences, under Award Number DE-FG02-03ER41270. 

\end{document}